\documentstyle[12pt,epsf,psfig]{article}

\newcommand{\gtrsim}{ \mathop{}_{\textstyle \sim}^{\textstyle >} }
\newcommand{\lesssim}{ \mathop{}_{\textstyle \sim}^{\textstyle <} }

\begin{document}
\baselineskip 0.7cm

\renewcommand{\thefootnote}{\fnsymbol{footnote}}
\setcounter{footnote}{1}

\begin{titlepage}
\begin{center}

\hfill    LBL-38099\\
\hfill    hep-ph/9512396\\

\vskip .5in

{\Large \bf The Muon Anomalous Magnetic Dipole Moment in the Minimal
Supersymmetric Standard Model}

\vskip .5in

{\large Takeo~Moroi}\footnote
{E-mail address : {\tt moroi@theor3.lbl.gov}}

\vskip .5in

{\it Theoretical Physics Group, Lawrence Berkeley National Laboratory\\
University of California, Berkeley, CA 94720, U.S.A.}

\end{center}

\vskip .5in

\begin{abstract}

The muon anomalous magnetic dipole moment (MDM) is calculated in the
framework of the minimal supersymmetric standard model (MSSM). In this
paper, we discuss how the muon MDM depends on the parameters in MSSM
in detail.  We show that the contribution of the superparticle-loop
becomes significant especially when $\tan\beta$ is large. Numerically,
it becomes $O(10^{-8}-10^{-9})$ in a wide parameter space, which is
within the reach of the new Brookhaven E821 experiment.

\end{abstract}

\vskip 1.5in

PACS numbers : 12.60.Jv, 13.40.Em, 14.60.Ef

\end{titlepage}

\renewcommand{\thefootnote}{\arabic{footnote}}
\setcounter{footnote}{0}

\section{Introduction}

Supersymmetry (SUSY)~\cite{NPB70-39} is one of the most attractive
candidates of the new physics beyond the standard model. In SUSY
models, quadratic divergences are automatically canceled out, and
hence SUSY may be regarded as a solution to the naturalness
problem~\cite{naturalness}. In addition, precision measurements of the
gauge coupling constants strongly suggest SUSY grand unified theory
(GUT)~\cite{unification}. Contrary to our theoretical interests,
however, evidences of SUSY (especially, superpartners) have not
discovered yet, and hence superpartners are fascinating targets of the
forthcoming high energy experiments like LEP~II, LHC and NLC.

Even if we do not have high energy colliders, we can constrain SUSY
models by using precision measurements in low energy experiments.
This is because superparticles contribute to low energy physics
through radiative corrections. Especially, superparticles are assumed
to have masses of the order of the electroweak scale, and hence their
loop effects may become comparable to those of $W^{\pm}$- or $Z$-boson
propagations.  Therefore, low energy precision experiments are also
very useful to obtain constraints on SUSY models.

One of the quantities which are measured in a great accuracy is the
muon anomalous magnetic dipole moment (MDM), $a_\mu\equiv
\frac{1}{2}(g-2)_\mu$. At present, the muon MDM is
measured to be~\cite{PDG}
\begin{eqnarray}
a_\mu^{\rm exp} = 1165923(8.4) \times 10^{-9}.
\label{g-2_exp}
\end{eqnarray}
On the other hand, the standard model prediction on $a_\mu$ is given
by~\cite{MDM_SM}
\begin{eqnarray}
a_\mu^{\rm SM} = 116591802(153) \times 10^{-11},
\label{g-2_sm}
\end{eqnarray}
which is completely consistent with experimental value. (For a review
of the calculation of $a_\mu^{\rm SM}$, see also
Ref.~\cite{kinoshita}.)

Because of the great accuracy of $a_\mu^{\rm exp}$ and $a_\mu^{\rm
SM}$ given above, we can derive a constraint on SUSY models from the
muon MDM. Furthermore, the new Brookhaven E821 experiment~\cite{E821}
is supposed to reduce the error of the experimental value of $a_\mu$
to $0.4\times 10^{-9}$, which is smaller than the present one by a
factor $\sim 20$. The accuracy of the Brookhaven E821 experiment is of
the order of the contribution of the $W^\pm$- and $Z$-boson loop,
which means we may have a chance to measure the SUSY contribution to
the muon MDM by that experiment.

In fact, there are several works in which the muon MDM is calculated
in the framework of SUSY models~\cite{MDM_SUSY,PRD49-366,9507386}.
Especially, Chattopadhyay and Nath recently pointed out that the muon
MDM is a powerful probe of the models based on supergravity if
$\tan\beta$ is large~\cite{9507386}. However, most of the recent works
assume the boundary conditions on the SUSY breaking parameters based
on the minimal supergravity, and/or radiative electroweak symmetry
breaking scenario, and hence it is quite unclear for us how the SUSY
contributions to the muon MDM depend on the parameters in MSSM. Thus,
the aim of this paper is to clarify it, and to investigate the
behavior of the muon MDM in the framework of MSSM. The mass matrices
and mixing angles among the superparticles have model dependence even
if we assume the boundary condition based on the minimal supergravity,
and hence we believe that it is important to analyze the muon MDM in a
more general framework of the SUSY standard model.

In this paper, we investigate the SUSY contribution to the muon MDM in
the framework of MSSM as a low energy effective theory of SUSY
GUT~\cite{SUSY_GUT}. The organization of this paper is as follows. In
the next section, we introduce a model we consider. In
Section~\ref{sec:analytic}, we show analytic forms of the SUSY
contribution to the muon MDM, $\Delta a_\mu^{\rm SUSY}$. In
Section~\ref{sec:behavior}, typical behavior of $\Delta a_\mu^{\rm
SUSY}$ is discussed. In Section~\ref{sec:numerical}, some numerical
results are shown.  Section~\ref{sec:discuss} is devoted to
discussion.

\section{Model}
\label{sec:model}

First of all, we would like to introduce a model we consider, {\it
i.e.} MSSM as a low energy effective theory of SUSY GUT. All the
fields we use in our analysis are
\begin{eqnarray}
l_L = ( \mu_L ~ \nu ),
~~~\mu_R^c, 
~~~H_1 = ( H_1^- ~ H_1^0 ),~~~
H_2 = \left( \begin{array}{c} H_2^+ \\ H_2^0 \end{array} \right),
\end{eqnarray}
where $l_L$ (${\bf 2^*}$, $-\frac{1}{2}$) and $\mu_R^c$ (${\bf 1}$,
$1$) are left- and right-handed leptons in the second generation,
while two Higgs doublets are
represented as $H_1$ (${\bf 2^*}$, $-\frac{1}{2}$) and $H_2$ (${\bf
2}$, $\frac{1}{2}$). (We denote the quantum numbers for the ${\rm
SU(2)_L}\times {\rm U(1)_Y}$ gauge group in the parenthesis.) The
Higgs doublets $H_1$ and $H_2$ are responsible for the electroweak
symmetry breaking, and hence their vacuum-expectation values are
constrained as $\langle H_1\rangle^2+\langle
H_2\rangle^2\simeq(174{\rm GeV})^2$ in order to give a correct value
of the Fermi constant. On the other hand, the ratio of the
vacuum-expectation values of two Higgs doublets is a free parameter in
MSSM, which we define $\tan\beta\equiv\langle H_2\rangle/\langle
H_1\rangle$.

Relevant part of the superpotential of MSSM is given by
\begin{eqnarray}
W = y_\mu \epsilon^{\alpha\beta} \mu_R^c l_{L,\alpha} H_{1,\beta}
+ \mu_H H_{1,\alpha} H_2^\alpha,
\end{eqnarray}
where $y_\mu$ is the Yukawa coupling constant of muon, $\mu_H$ the
SUSY invariant Higgs mass and $\epsilon^{\alpha\beta}$ the
anti-symmetric tensor with $\epsilon^{12}=1$. Using the superpotential
given above, $F$-term contribution to the lagrangian is obtained as
\begin{eqnarray}
{\cal L}_F = - \int d^2\theta W + h.c.
\end{eqnarray}
Furthermore, soft SUSY breaking terms are given by
\begin{eqnarray}
{\cal L}_{\rm soft} &=& -m_L^2 \tilde{l}_{L}^* \tilde{l}_{L}
- m_R^2 \tilde{\mu}_R^{c*} \tilde{\mu}_{R}^c
- (A_\mu 
\epsilon^{\alpha\beta} \tilde{\mu}_R^c \tilde{l}_{L,\alpha} H_{2,\beta}
+ h.c.)
\nonumber \\ &&
- \frac{1}{2} (m_{G2} \tilde{W}\tilde{W} 
+ m_{G1} \tilde{B}\tilde{B} + h.c.).
\label{L_soft}
\end{eqnarray}
Here, $\tilde{l}_L$, $\tilde{\mu}_R^c$, $\tilde{W}$ and $\tilde{B}$
represent left- and right-handed sleptons in second generation, and
gauginos for ${\rm SU(2)_L}$ and ${\rm U(1)_Y}$ gauge group,
respectively. Gaugino masses $m_{G1}$ and $m_{G2}$ are related by the
GUT relation;
\begin{eqnarray}
\frac{m_{G2}}{g_2^2} = \frac{3}{5} \frac{m_{G1}}{g_1^2},
\label{GUT_relation}
\end{eqnarray}
where $g_1$ and $g_2$ are the gauge coupling constant of ${\rm
SU(2)_L}$ and ${\rm U(1)_Y}$ gauge group, respectively.\footnote
{The GUT relation given in eq.(\ref{GUT_relation}) holds in general if
the gauge groups are unified in a larger group~\cite{PLB324-52}.
Therefore, we are not depending on specific model of GUT.}

Here, we should comment on a flavor mixing in slepton mass matrices.
If there is large flavor mixings in the slepton mass matrices, all
the sleptons contribute to the muon MDM. However, flavor mixing in the
slepton mass matrices may be dangerous, since it induces lepton
flavor violation processes such as $\mu\rightarrow e\gamma$,
$\tau\rightarrow\mu\gamma$ and so on. Especially, the mixing among
first and second generation is severely constrained from
$\mu\rightarrow e\gamma$ especially when $\tan\beta$ is
large~\cite{lfv}. On the other hand, the constraint on the mixing of
the second and third generation is not so stringent. In this paper,
for simplicity, we assume that the flavor mixing in the slepton mass
matrix is not so large, and that it does not affect the following
arguments. A comment on the case with the flavor mixing is given in
Section~\ref{sec:discuss}.

Once we have the MSSM lagrangian, we can obtain mass eigenvalues and
mixing matrices of the superparticles. The mass matrix for the smuon
field is given by
\begin{eqnarray}
M_{\tilde{\mu}}^2 =
\left( \begin{array}{cc}
m_{\tilde{\mu}L}^2 & m_{LR}^2 \\
m_{LR}^2 & m_{\tilde{\mu}R}^2 
\end{array} \right),
\label{M_smu}
\end{eqnarray}
where 
\begin{eqnarray}
m_{\tilde{\mu}L}^2 &=& m_L^2 + m_Z^2 \cos 2\beta 
\left( \sin^2\theta_W - \frac{1}{2} \right),
\\
m_{\tilde{\mu}R}^2 &=& m_R^2 - m_Z^2 \cos 2\beta 
\sin^2\theta_W,
\\
m_{LR}^2 &=&
y_\mu \mu_H \langle H_2\rangle + A_\mu \langle H_1\rangle.
\label{m_LR}
\end{eqnarray}
The mass matrix $M_{\tilde{\mu}}^2$ can be diagonalized by using an
unitary matrix $U_{\tilde{\mu}}$ as
\begin{eqnarray}
(U_{\tilde{\mu}}^\dagger M_{\tilde{\mu}}^2U_{\tilde{\mu}})_{AB} =
m_{\tilde{\mu}A}^2 \delta_{AB} ~~~ (A,B = 1,2),
\end{eqnarray}
where $m_{\tilde{\mu}A}$ is the mass eigenvalue of the smuon. Notice
that, in our case, off-diagonal element of the mass matrix given in
eq.(\ref{M_smu}) is substantially smaller than the diagonal elements,
and hence $m_{\tilde{\mu}L}$ and $m_{\tilde{\mu}R}$ almost correspond
to the mass eigenvalues. The mass of the sneutrino, $m_{\tilde{\nu}}$,
is also easily obtained as
\begin{eqnarray}
m_{\tilde{\nu}}^2 = m_L^2 + \frac{1}{2} m_Z^2 \cos 2\beta.
\end{eqnarray}

Next, we derive the mass matrices for neutralinos and charginos.  For
neutralinos, the mass terms are given by
\begin{eqnarray}
{\cal L}_{\chi^0} &=&
-\frac{1}{2}
(i\tilde{B}~i\tilde{W}^3~\tilde{H}_1^0~\tilde{H}_2^0)
\nonumber \\ &&
\left( \begin{array}{cccc}
-m_{G1} & 0 & 
-\frac{1}{\sqrt{2}}g_1\langle H_1\rangle &
\frac{1}{\sqrt{2}}g_1\langle H_2\rangle \\
0 & -m_{G2} &
\frac{1}{\sqrt{2}}g_2\langle H_1\rangle &
- \frac{1}{\sqrt{2}}g_2\langle H_2\rangle \\
-\frac{1}{\sqrt{2}}g_1\langle H_1\rangle &
\frac{1}{\sqrt{2}}g_2\langle H_1\rangle &
0 & \mu_H \\
\frac{1}{\sqrt{2}}g_1\langle H_2\rangle &
- \frac{1}{\sqrt{2}}g_2\langle H_2\rangle &
\mu_H & 0
\end{array}\right)
\left( \begin{array}{c}
i\tilde{B} \\ i\tilde{W}^3 \\ \tilde{H}_1^0 \\ \tilde{H}_2^0
\end{array}\right)
\nonumber \\ &&
+ h.c.,
\label{mass_chi0}
\end{eqnarray}
where $\tilde{H}_1$ and $\tilde{H}_2$ represent the higgsino field.
Then, we can find an unitary matrix $U_{\chi^0}$ which diagonalize the
mass matrix given above. Denoting the mass matrix given in
eq.(\ref{mass_chi0}) as $M_{\chi^0}$, mass eigenvalues $m_{\chi^0X}$
for the neutralino $\chi^0_X$ is given by
\begin{eqnarray}
(U_{\chi^0}^\dagger M_{\chi^0} U_{\chi^0})_{XY}
= m_{\chi^0X} \delta_{XY} ~~~ (X,Y = 1 - 4),
\end{eqnarray}
where $m_{\chi^0X}\leq m_{\chi^0Y}$ if $X<Y$. 

Similarly, mass terms for the charginos are given by
\begin{eqnarray}
{\cal L}_{\chi^\pm} = 
- (\tilde{W}^+~\tilde{H}_2^+)
\left( \begin{array}{cc}
-m_{G2} &
g_2\langle H_1\rangle \\
- g_2\langle H_2\rangle &
\mu_H
\end{array}\right)
\left( \begin{array}{c}
\tilde{W}^- \\ \tilde{H}_1^-
\end{array}\right) + h.c.~,
\label{mass_chi+-}
\end{eqnarray}
with $\tilde{W}^\pm\equiv -\frac{i}{\sqrt{2}} (\tilde{W}^1\mp
i\tilde{W}^2)$. The mass matrix given in eq.(\ref{mass_chi+-}), which
we denote $M_{\chi^\pm}$, can be diagonalized by using two unitary
matrices, $U_{\chi^+}$ and $U_{\chi^-}$;
\begin{eqnarray}
(U_{\chi^+}^\dagger M_{\chi^\pm} U_{\chi^-})_{XY}
= m_{\chi^\pm X} \delta_{XY} ~~~ (X,Y = 1,2),
\end{eqnarray}
where $m_{\chi^\pm X}$ represents the mass eigenvalue of the chargino
field.

With the coupling constants and mixing matrices given above, we can
write down muon-neutralino-smuon and muon-chargino-sneutrino vertices.
Denoting the mass eigenstates of the smuon, neutralino and chargino as
$\tilde{\mu}_A$, $\chi^0_X$ and $\chi^\pm_X$, respectively, the
interaction terms are given by
\begin{eqnarray}
{\cal L}_{\rm int} =
\sum_{AX} \bar{\mu} (N_{AX}^{L}P_L + N_{AX}^RP_R) 
\chi^0_X \tilde{\mu}_A
+ \sum_{X} \bar{\mu} (C_{X}^{L}P_L + C_{X}^RP_R) 
\chi^\pm_X \tilde{\nu} + h.c.~,
\label{L_int}
\end{eqnarray}
where $P_L =\frac{1}{2}(1-\gamma_5)$, $P_R =\frac{1}{2}(1+\gamma_5)$
and
\begin{eqnarray}
N_{AX}^L &=& -y_\mu (U_{\chi^0})_{3X} (U_{\tilde{\mu}})_{LA}
-\sqrt{2} g_1 (U_{\chi^0})_{1X} (U_{\tilde{\mu}})_{RA},
\\
N_{AX}^R &=& -y_\mu (U_{\chi^0})_{3X} (U_{\tilde{\mu}})_{RA}
+\frac{1}{\sqrt{2}} g_2 (U_{\chi^0})_{2X} (U_{\tilde{\mu}})_{LA}
+\frac{1}{\sqrt{2}} g_1 (U_{\chi^0})_{1X} (U_{\tilde{\mu}})_{LA},
\\
C_{X}^L &=& y_\mu (U_{\chi^-})_{2X},
\\
C_{X}^R &=& -g_2 (U_{\chi^+})_{1X}.
\end{eqnarray}
By using the interaction terms given in eq.(\ref{L_int}), we calculate
the SUSY contribution to the muon MDM.

\section{Analytic formulae}
\label{sec:analytic}

Now, we are in position to calculate the SUSY contribution to the muon
MDM. What we have to calculate is the ``magnetic moment type''
operator, which is given by
\begin{eqnarray}
{\cal L}_{\rm MDM} = \frac{e}{4m_\mu} F_2
\bar{\mu} \sigma_{\rho\lambda} \mu F^{\rho\lambda}.
\label{mag_mom}
\end{eqnarray}
Here, $e$ is the electric charge, $m_\mu$ the muon mass,
$\sigma_{\rho\lambda}=\frac{i}{2}[\gamma_\rho ,\gamma_\lambda]$,
$F_{\rho\lambda}$ the field strength of the photon field and $F_2$ the
magnetic form factor. The muon anomalous magnetic moment, $a_\mu$, is
related to $F_2$ as
\begin{eqnarray}
a_\mu = F_2.
\end{eqnarray}
Thus, by calculating magnetic form factor in the framework of MSSM, we
can have SUSY contribution to the muon MDM.

In SUSY model, there are essentially two types of diagrams which
contribute to $a_\mu$, {\it i.e.} one is the neutralino
($\chi^0$)-smuon ($\tilde{\mu}$) loop diagram (Fig.~\ref{fig:feyn_1}a)
and the other is the chargino ($\chi^{\pm}$)-sneutrino ($\tilde{\nu}$)
loop diagram (Fig.~\ref{fig:feyn_1}b);
\begin{eqnarray}
\Delta a_\mu^{\rm SUSY} = 
\Delta a_\mu^{\chi^0\tilde{\mu}} + \Delta a_\mu^{\chi^\pm\tilde{\nu}}.
\end{eqnarray}
Here, contribution from the $\chi^0$-$\tilde{\mu}$ diagram, $\Delta
a_\mu^{\chi^0\tilde{\mu}}$, is 
\begin{eqnarray}
\Delta a_\mu^{\chi^0\tilde{\mu}} &=& 
m_\mu \sum_{AX} \Big\{-m_\mu 
(N^L_{AX} N^L_{AX} + N^R_{AX} N^R_{AX}) m_{\tilde{\mu}A}^2 
J_5(m_{\chi^0X}^2,m_{\tilde{\mu}A}^2,m_{\tilde{\mu}A}^2,
m_{\tilde{\mu}A}^2,m_{\tilde{\mu}A}^2)
\nonumber \\ &&
+ m_{\chi^0X} N^L_{AX} N^R_{AX} 
J_4(m_{\chi^0X}^2,m_{\chi^0X}^2,m_{\tilde{\mu}A}^2,m_{\tilde{\mu}A}^2) \Big\}
\nonumber \\ &=&
\frac{1}{16\pi^2} m_\mu \sum_{AX} 
\Bigg\{-\frac{m_\mu}{6m_{\tilde{\mu}A}^2(1-x_{AX})^4}
(N^L_{AX} N^L_{AX} + N^R_{AX} N^R_{AX})
\nonumber \\ &&
\times (1-6x_{AX}+3x_{AX}^2+2x_{AX}^3-6x_{AX}^2\ln x_{AX})
\nonumber \\ &&
- \frac{m_{\chi^0X}}{m_{\tilde{\mu}A}^2(1-x_{AX})^3} N^L_{AX} N^R_{AX}
(1-x_{AX}^2+2x_{AX}\ln x_{AX})
\Bigg\},
\label{g-2_nAX}
\end{eqnarray}
where we are using mass eigenstate basis of $\chi^0$ and $\tilde{\mu}$
(and that of $\chi^\pm$ in deriving eq.(\ref{g-2_cAX})).  Here,
$x_{AX}=m_{\chi^0X}^2/m_{\tilde{\mu}A}^2$, and we define the functions
$I_N$ and $J_N$ as
\begin{eqnarray}
I_N(m_1^2,\cdots,m_N^2) &=&
\int\frac{d^4k}{(2\pi)^4i}
\frac{1}{(k^2-m_1^2)\cdots (k^2-m_N^2)},
\label{I_N} \\
J_N(m_1^2,\cdots,m_N^2) &=&
\int\frac{d^4k}{(2\pi)^4i}
\frac{k^2}{(k^2-m_1^2)\cdots (k^2-m_N^2)}.
\label{J_N}
\end{eqnarray}
Some useful formulae concerning the functions $I_N$ and $J_N$ are
shown in Appendix~\ref{ap:i&j}.  Contribution from the
$\chi^\pm$-$\tilde{\nu}$ loop diagram is also easily calculated, and
the result is given by
\begin{eqnarray}
\Delta a_\mu^{\chi^\pm\tilde{\nu}} &=& 
m_\mu \sum_{X}\Big[ m_\mu 
(C^L_{X} C^L_{X} + C^R_{X} C^R_{X})
\{ J_4(m_{\chi^\pm X}^2,m_{\chi^\pm X}^2,m_{\chi^\pm X}^2,m_{\chi^\pm X}^2) 
\nonumber \\ &&
+ m_{\tilde{\nu}}^2 J_5(m_{\chi^\pm X}^2,m_{\chi^\pm X}^2,m_{\chi^\pm X}^2
,m_{\chi^\pm X}^2,m_{\tilde{\nu}}^2)
- J_4(m_{\chi^\pm X}^2,m_{\chi^\pm X}^2,m_{\chi^\pm X}^2
,m_{\tilde{\nu}}^2) \}
\nonumber \\ &&
-2m_X C^L_{X} C^R_{X}
J_4 (m_{\chi^\pm X}^2,m_{\chi^\pm X}^2,m_{\chi^\pm X}^2
,m_{\tilde{\nu}}^2) \Big]
\nonumber \\ &=&
\frac{1}{16\pi^2} m_\mu \sum_{X} 
\Bigg\{\frac{m_\mu}{3m_{\tilde{\nu}}^2(1-x_{X})^4}
(C^L_{X} C^L_{X} + C^R_{X} C^R_{X})
\nonumber \\ &&
\times \left(
1 + \frac{3}{2} x_{X} - 3x_{X}^2 + \frac{1}{2}x_{X}^3
+ 3x_{X}\ln x_{X} \right)
\nonumber \\ &&
- \frac{3m_{\chi^\pm X}}{m_{\tilde{\nu}}^2(1-x_{X})^3} C^L_{X} C^R_{X}
\left( 1 - \frac{4}{3}x_{X} + \frac{1}{3}x_{X}^2
+ \frac{2}{3}\ln x_{X} \right)
\Bigg\},
\label{g-2_cAX}
\end{eqnarray}
where $x_X =m_{\chi^\pm X}^2/m_{\tilde{\nu}}^2$.

\section{Behavior of the SUSY contribution to the muon MDM}
\label{sec:behavior}

Before evaluating the SUSY contribution to the muon MDM numerically,
we would like to discuss the behavior of $\Delta a_\mu^{\rm SUSY}$,
especially in large $\tan\beta$ case.  As we will soon see, $|\Delta
a_\mu^{\rm SUSY}|$ becomes large as $\tan\beta$ increases. Thus, the
discussion about the large $\tan\beta$ case will be helpful for us to
understand the behavior of $\Delta a_\mu^{\rm SUSY}$.

For this purpose, it is more convenient to use the mass insertion
method to calculate the penguin diagrams rather than working in the
mass eigenstate basis of the superparticles which is used in the
previous section. In the case where $\tan\beta$ is large, five
diagrams dominantly contribute to $\Delta a_\mu^{\rm SUSY}$, which are
shown in Fig.~\ref{fig:feyn_2}. Their contributions are given by
\begin{eqnarray}
\Delta a^{\rm N1}_\mu &=&
g_1^2 m_\mu^2 m_{G1} \mu_H \tan\beta
\nonumber \\ &&
\times \Big\{
J_5(m_{G1}^2,m_{G1}^2,m_{\tilde{\mu}L}^2,m_{\tilde{\mu}R}^2,m_{\tilde{\mu}R}^2)
+ J_5(m_{G1}^2,m_{G1}^2,m_{\tilde{\mu}L}^2,m_{\tilde{\mu}L}^2,m_{\tilde{\mu}R}^2)
\Big\},
\label{dg_n1}
\\
\Delta a^{\rm N2}_\mu &=&
- g_1^2 m_\mu^2 m_{G1} \mu_H \tan\beta
\nonumber \\ &&
\times \Big\{
J_5(m_{G1}^2,m_{G1}^2,\mu_H^2,m_{\tilde{\mu}R}^2,m_{\tilde{\mu}R}^2)
+ J_5(m_{G1}^2,\mu_H^2,\mu_H^2,m_{\tilde{\mu}R}^2,m_{\tilde{\mu}R}^2)
\Big\},
\label{dg_n2}
\\
\Delta a^{\rm N3}_\mu &=&
\frac{1}{2} g_1^2 m_\mu^2 m_{G1} \mu_H \tan\beta
\nonumber \\ &&
\times \Big\{
J_5(m_{G1}^2,m_{G1}^2,\mu_H^2,m_{\tilde{\mu}L}^2,m_{\tilde{\mu}L}^2)
+ J_5(m_{G1}^2,\mu_H^2,\mu_H^2,m_{\tilde{\mu}L}^2,m_{\tilde{\mu}L}^2)
\Big\},
\label{dg_n3}
\\
\Delta a^{\rm N4}_\mu &=&
- \frac{1}{2} g_2^2 m_\mu^2 m_{G2} \mu_H \tan\beta
\nonumber \\ &&
\times \Big\{
J_5(m_{G2}^2,m_{G2}^2,\mu_H^2,m_{\tilde{\mu}L}^2,m_{\tilde{\mu}L}^2)
+ J_5(m_{G2}^2,\mu_H^2,\mu_H^2,m_{\tilde{\mu}L}^2,m_{\tilde{\mu}L}^2)
\Big\},
\label{dg_n4}
\\
\Delta a^{\rm C}_\mu &=&
g_2^2 m_\mu^2 m_{G2} \mu_H \tan\beta
\nonumber \\ &&
\times \Big\{
2 I_4(m_{G2}^2,m_{G2}^2,\mu_H^2,m_{\tilde{\nu}}^2)
- J_5(m_{G2}^2,m_{G2}^2,\mu_H^2,m_{\tilde{\nu}}^2,m_{\tilde{\nu}}^2)
\nonumber \\ &&
+ 2 I_4(m_{G2}^2,\mu_H^2,\mu_H^2,m_{\tilde{\nu}}^2)
- J_5(m_{G2}^2,\mu_H^2,\mu_H^2,m_{\tilde{\nu}}^2,m_{\tilde{\nu}}^2)
\Big\}.
\label{dg_c}
\end{eqnarray}
Here, eqs.(\ref{dg_n1}) -- (\ref{dg_n4}) are $\chi^0$-$\tilde{\mu}$
loop contributions, while eq.(\ref{dg_c}) represents the
$\chi^\pm$-$\tilde{\nu}$ loop one. By using these expressions, the
SUSY contribution to the muon MDM is approximately given by
\begin{eqnarray}
\Delta a_\mu^{\chi^0\tilde{\mu}} &\simeq&
\Delta a^{\rm N1}_\mu + \Delta a^{\rm N2}_\mu + 
\Delta a^{\rm N3}_\mu + \Delta a^{\rm N4}_\mu,
\\
\Delta a_\mu^{\chi^\pm\tilde{\nu}} &\simeq&
\Delta a^{\rm C}_\mu.
\end{eqnarray}
Notice that the SUSY contribution to the muon MDM given in
eqs.(\ref{dg_n1}) -- (\ref{dg_c}) approximately correspond to the
terms which are proportional to $N^LN^R$ or $C^LC^R$ ({\it i.e.} terms
which have a chirality flip in internal fermion line) in the exact
formulae given in eqs.(\ref{g-2_nAX}) and (\ref{g-2_cAX}).

The first thing we can learn from the above expressions is that all
the terms given in eqs.(\ref{dg_n1}) -- (\ref{dg_c}) are proportional
to $\tan\beta$~\cite{PRD49-366,9507386}. This is due to the fact that
the chirality is flipped not by hitting the mass of the external muon
but by directly hitting the Yukawa coupling. This mechanism also
occurs in the case of the lepton flavor violations~\cite{lfv}.  Thus,
$|\Delta a_\mu^{\rm SUSY}|$ becomes large as $\tan\beta$ increases,
and we obtain severer constraint on the parameter space as $\tan\beta$
gets larger.

The second point we should mention is that the relation between the
sign of $\Delta a_\mu^{\rm SUSY}$ and those of parameters in MSSM. The
dominant SUSY contribution given in eqs.(\ref{dg_n1}) -- (\ref{dg_c})
are all proportional to $m_{G}\mu_H\tan\beta$ (with
$m_{G}=m_{G1},m_{G2}$ being gaugino mass). Thus, if we change the sign
of this combination, $\Delta a_\mu^{\rm SUSY}$ also changes its sign.
Furthermore, in the case where we assume GUT relation on the gaugino
masses, we checked that $\Delta a_\mu^{\rm N1}$ or $\Delta a_\mu^{\rm
C}$ dominates over other terms ($\Delta a_\mu^{\rm N2}$, $\Delta
a_\mu^{\rm N3}$, $\Delta a_\mu^{\rm N4}$) in most of the parameter
space.  Here, both $\Delta a_\mu^{\rm N1}$ and $\Delta a_\mu^{\rm C}$
have the same sign as the combination $m_{G}\mu_H\tan\beta$.
Therefore, $\Delta a_\mu^{\rm SUSY}$ becomes positive (negative) when
the sign of the combination $m_{G}\mu_H\tan\beta$ is positive
(negative).\footnote
{If $m_{G}$ or $\mu_H$ is small, this relation does not hold. This is
mainly because that the mass insertion method breaks down in such a
case.  Furthermore, in such a case, we cannot ignore $\Delta
a_\mu^{\rm N2}$ or terms which is not proportional to $\tan\beta$
({\it i.e.} terms which are proportional to $N^LN^L$, $N^RN^R$,
$C^LC^L$ and $C^RC^R$ in the exact formula given in
eqs.(\ref{g-2_nAX}) and (\ref{g-2_cAX})). In that case, the sign of
$m_{G}\mu_H\tan\beta$ is not directly related to that of $\Delta
a_\mu^{\rm SUSY}$.}
In the next section, we will see that this relation really holds as a
result of numerical calculations.

Furthermore, we comment here that the contribution of
$\chi^\pm$-$\tilde{\nu}$ loop diagram dominates over that of the
$\chi^0$-$\tilde{\mu}$ loop ones if all the masses of the
superparticles are almost degenerate.  For example, let us consider
the extreme case where all the masses for the superparticles
($m_{G1}$, $m_{G2}$, $\mu_H$, $m_{\tilde{\mu}L}$, $m_{\tilde{\mu}R}$,
$m_{\tilde{\nu}}$) are the same. Denoting the masses of the
superparticles $m_{\rm SUSY}$, contributions of the
$\chi^0$-$\tilde{\mu}$ and $\chi^\pm$-$\tilde{\nu}$ loop diagrams to
the muon MDM is given by
\begin{eqnarray}
\Delta a^{\chi^0\tilde{\mu}}_\mu &\simeq& 
\Delta a^{\rm N1}_\mu + \Delta a^{\rm N2}_\mu + \Delta a^{\rm N3}_\mu
+ \Delta a^{\rm N4}_\mu
\nonumber \\ &=&
\frac{1}{192\pi^2} \frac{m_\mu^2}{m_{\rm SUSY}^2}(g_1^2 - g_2^2) 
\tan\beta,
\\
\Delta a^{\chi^\pm\tilde{\nu}}_\mu &\simeq& \Delta a^{\rm C}_\mu
\nonumber \\ &=&
\frac{1}{32\pi^2} \frac{m_\mu^2}{m_{\rm SUSY}^2}g_2^2
\tan\beta.
\end{eqnarray}
From the above expressions, we can see that the
$\chi^\pm$-$\tilde{\nu}$ loop contribution is substantially larger
than that of $\chi^0$-$\tilde{\mu}$ loop. Thus,
$\chi^\pm$-$\tilde{\nu}$ loop gives dominant contribution, as in the
case of minimal SUSY GUT based on the minimal
supergravity~\cite{9507386}.  However, we should note here that the
$\chi^\pm$-$\tilde{\nu}$ loop dominance does not hold in general. In
the next section, we will see the SUSY contribution to $\Delta
a_\mu^{\rm SUSY}$ significantly depends on the right-handed smuon mass
$m_{\tilde{\mu}R}$ in certain parameter regions.

\section{Numerical Results}
\label{sec:numerical}

In this section, we numerically estimate $\Delta a_\mu^{\rm SUSY}$ by
using eqs.(\ref{g-2_nAX}) and (\ref{g-2_cAX}). As we mentioned
before, there are essentially six parameters on which $\Delta
a_\mu^{\rm SUSY}$ depends, {\it i.e.} SU(2) gaugino mass
$m_{G2}$,\footnote
{Gaugino mass for ${\rm U(1)_Y}$ gauge group is determined by the GUT
relation~(\ref{GUT_relation}).}
left- and right-handed smuon masses $m_{\tilde{\mu}L}$ and
$m_{\tilde{\mu}R}$ (which essentially correspond to the soft SUSY
breaking parameters $m_L^2$ and $m_R^2$, respectively), SUSY invariant
Higgs mass $\mu_H$, ratio of the vacuum-expectation values of the two
Higgs doublets, $\tan\beta =\langle H_2\rangle /\langle H_1\rangle$,
and SUSY breaking $A$-parameter for the smuon, $A_\mu$. However,
especially in the large $\tan\beta$ region where SUSY contribution to
$\Delta a_\mu^{\rm SUSY}$ may become significantly large, $\Delta
a_\mu^{\rm SUSY}$ is not sensitive to $A_\mu$ if $A_\mu\sim
O(y_\mu\mu_H)$. This is because $A_\mu$ always appears in expressions
in the combination of $(A_\mu +y_\mu\mu_H\tan\beta)$, as shown in
eq.(\ref{m_LR}).  Therefore, in our analysis, we take $A_\mu
=0$.\footnote
{The supergravity model suggests $A_\mu\sim O(y_\mu
m_{\tilde{\mu}})$~\cite{PRepC110-1}. Furthermore, it was pointed out
that some unwanted minimum appears in the potential of the smuon when
$|A_\mu| >O(1)\times y_\mu m_{\tilde{\mu}}$~\cite{NPB237-307}, which
may cause cosmological difficulties. We checked that the results
are almost unchanged even if we take $A_\mu =3y_\mu m_{\tilde{\mu}L}$.}
Then, we take the other five parameters as free parameters and
calculate the SUSY contribution to the muon MDM for a given set of
parameters.

First, we show the SUSY contribution to the muon MDM for fixed values
of $m_{\tilde{\mu}R}$ and $m_{\tilde{\mu}L}$ in the $\mu_H$-$m_{G2}$
plane. In Fig.~\ref{fig:r100gev}, we plotted the results for
$m_{\tilde{\mu}R}=100{\rm GeV}$ and $\tan\beta =30$. Here, the
left-handed smuon mass is taken to be 100GeV
(Fig.~\ref{fig:r100gev}a), 300GeV (Fig.~\ref{fig:r100gev}b) and 500GeV
(Fig.~\ref{fig:r100gev}c). The results for the cases of
$m_{\tilde{\mu}R}=300{\rm GeV}$ and $m_{\tilde{\mu}R}=1{\rm TeV}$ are
also shown in Fig.~\ref{fig:r300gev} and Fig.~\ref{fig:r1tev},
respectively.  As we can see, if we take a smaller value of
$m_{\tilde{\mu}R}$, the SUSY contribution to the muon MDM is enhanced
in the large $\mu_H$ region.  This can be easily understood if we
think of the fact that $\Delta a^{\rm N1}_{\mu}$ gives a large
contribution in such a parameter region.

Furthermore, by choosing the right-handed smuon mass
$m_{\tilde{\mu}R}$ so that $|\Delta a_\mu^{\rm SUSY}|$ is minimized,
we obtain the lowerbound on the SUSY contribution to the muon MDM. The
results are shown in Fig.~\ref{fig:min}. Here, we assume $45{\rm
GeV}\leq m_{\tilde{\mu}R}\leq 1{\rm TeV}$. The lowerbound is obtained
from the negative search for the smuon~\cite{PDG}, while the
upperbound is due to the naturalness point of view. In fact, the
results are insensitive to the upperbound if we take the upperbound
larger than about 1TeV, since the effects of the right-handed smuon
decouple when we take $m_{\tilde{\mu}R}\rightarrow\infty$.

Here, we would like to discuss the behavior of $\Delta a_\mu^{\rm
SUSY}$ shown in Fig.~\ref{fig:min}.  As can be seen, $\Delta
a_\mu^{\rm SUSY}$ changes its behavior at around $|\mu_H|\sim
m_{\tilde{\mu}L}$. This can be understood in a following way. In the
case of $|\mu_H|\sim m_{\tilde{\mu}L}$, $\Delta a_\mu^{\rm N1}$ and
$\Delta a_\mu^{\rm N2}$ almost cancels out and $\Delta a_\mu^{\rm
SUSY}$ becomes insensitive to $m_{\tilde{\mu}R}$.  In the case of
$|\mu_H|\gtrsim m_{\tilde{\mu}L}$, $m_{\tilde{\mu}R}$-dependence of
$\Delta a_\mu^{\rm SUSY}$ is almost determined by that of $\Delta
a_\mu^{\rm N1}$. Then, $|\Delta a_\mu^{\rm SUSY}|$ becomes smaller as
$m_{\tilde{\mu}R}$ becomes larger. On the other hand, if
$|\mu_H|\lesssim m_{\tilde{\mu}L}$, $\Delta a_\mu^{\rm N2}$ determines
the $m_{\tilde{\mu}R}$-dependence of $\Delta a_\mu^{\rm SUSY}$. The
important point is that the sign of $\Delta a_\mu^{\rm N2}$ is
opposite to that of $\Delta a_\mu^{\rm C}$ which gives the dominant
contribution.  Thus, $|\Delta a_\mu^{\rm SUSY}|$ gets smaller as
$m_{\tilde{\mu}R}$ decreases. In summary, in the case of
$|\mu_H|\gtrsim m_{\tilde{\mu}L}$, $|\Delta a_\mu^{\rm SUSY}|$
increases as $m_{\tilde{\mu}R}$ decreases, while in the case of
$|\mu_H|\lesssim m_{\tilde{\mu}L}$, $|\Delta a_\mu^{\rm SUSY}|$
decreases as $m_{\tilde{\mu}R}$ gets smaller.

Notice that some regions of the $\mu_H$-$m_{G2}$ plane are excluded by
the negative search for signals of neutralinos or
charginos~\cite{lep,PLB350-109}. In Fig.~\ref{fig:lep}, we show the
excluded region for $\tan\beta =30$, {\it i.e.} for large
$\tan\beta$ case.\footnote
{If $\tan\beta$ is fairly large ($\tan\beta\gtrsim 5$), mass matrices
of the charginos and neutralinos become almost independent of
$\tan\beta$. In this case, the constraint is insensitive to
$\tan\beta$. We would like to note here that, if $\tan\beta$ is not so
large, in our convention, the constraint becomes severer for the case
of $\mu_Hm_{G2}>0$ rather than $\mu_Hm_{G2}<0$, as shown in
Refs.~\cite{lep,PLB350-109}.}
Here, we adopt the following constraints~\cite{PLB350-109};
\begin{eqnarray}
\Delta\Gamma_Z &<& 23.1{\rm MeV},
\\
\Delta\Gamma_{\rm inv} &<& 8.4{\rm MeV},
\\
Br(Z\rightarrow \chi^0_1\chi^0_2) &<& 5\times 10^{-5},
\\
Br(Z\rightarrow \chi^0_2\chi^0_2) &<& 5\times 10^{-5},
\end{eqnarray}
where $\Delta\Gamma_Z$ is the partial width of $Z$-boson decaying into
charginos or neutralinos, while $\Delta\Gamma_{\rm
inv}=\Gamma_Z(Z\rightarrow \chi^0_1\chi^0_1)$ represents the
neutralino contribution to the invisible width. For 
the constraint on the chargino mass, we
consider several cases where the lowerbound on the chargino mass is
given by 45GeV (LEP), 90GeV (LEP~II), and 250GeV (NLC with
$\sqrt{s}=500{\rm GeV}$). Comparing Fig.~\ref{fig:lep} with
Fig.~\ref{fig:r100gev} -- Fig.~\ref{fig:min}, we can see that the
muon MDM has a better sensitivity to MSSM than colliders in some
parameter space.

Remember that the SUSY contribution to the muon MDM is approximately
proportional to $\tan\beta$. Therefore, even for the case of
$\tan\beta\neq 30$, we can read off the approximate value of $\Delta
a_\mu^{\rm SUSY}$ from Fig.~\ref{fig:r100gev} -- Fig.~\ref{fig:min}.
For example, the contours for $\Delta a_\mu^{\rm SUSY}=2\times
10^{-9}$ in these figures correspond to $\Delta a_\mu^{\rm SUSY}\simeq
4\times 10^{-9}$ for the case of $\tan\beta =60$.

If the new Brookhaven E821 experiment measures the muon MDM with the
accuracy of their proposal, it will give a great impact on MSSM. In a
large parameter space, $|\Delta a_\mu^{\rm SUSY}|$ becomes
$O(10^{-9})$, which is within the reach of the new Brookhaven E821
experiment. Furthermore, the theoretical uncertainty, which is almost
originate to the hadronic uncertainty, is also expected to be
decreased due to better measurements of the cross section of
$e^++e^-\rightarrow {\rm hadrons}$ at low energies. Thus, the muon MDM
should be regarded as a good probe of MSSM.  In particular, the
Brookhaven E821 may be able to see the signal of MSSM even in the case
where we cannot find any superparticle by NLC with $\sqrt{s}=500{\rm
GeV}$.

The SUSY contribution should be compared with the present constraints
on the muon MDM from experiment and theoretical calculations, which
are given in eqs.(\ref{g-2_exp}) and (\ref{g-2_sm}).  Combining them,
we obtain a constraint on the SUSY contribution to the muon MDM,
$\Delta a_\mu^{\rm SUSY}$, which is given by
\begin{eqnarray}
-9.0\times 10^{-9} \leq \Delta a_\mu^{\rm SUSY}
\leq 19.0\times 10^{-9} ~~~({\rm 90\%~C.L.}).
\label{constraint_dg}
\end{eqnarray}
In Fig.~\ref{fig:now}, we show the contour of $\tan\beta$ which gives
the threshold value of the present constraint on $\Delta a_\mu^{\rm
SUSY}$ given above ({\it i.e.} $\Delta a_\mu^{\rm SUSY} = -9.0\times
10^{-9}$ and $\Delta a_\mu^{\rm SUSY} = 19.0\times 10^{-9}$).  Here,
we choose $m_{\tilde{\mu}R}$ so that $|\Delta_\mu^{\rm SUSY}-5.0\times
10^{-9}|$ is minimized (where $5.0\times 10^{-9}$ is the center value
of the constraint (\ref{constraint_dg})). Thus, Fig.~\ref{fig:now}
should be regarded as a constraint on the $\mu_H$-$m_{G2}$ plane for a
fixed values of $m_{\tilde{\mu}L}$ and $\tan\beta$.  Notice that if we
assume a larger value of $\tan\beta$, SUSY contribution exceeds the
present limit on the muon MDM in wider regions.

Before closing this section, we point out the fact that the contour in
Fig.~\ref{fig:now} is not symmetric under $\mu_H\rightarrow -\mu_H$.
This is because the center value of the constraint given in inequality
(\ref{constraint_dg}) is $5.0\times 10^{-9}$, which deviates from
zero. Therefore, constraint (\ref{constraint_dg}) prefers positive
value of $\Delta a_\mu^{\rm SUSY}$, and hence we have severer
constraint for $\mu_H <0$.

\section{Discussion}
\label{sec:discuss}

In this paper, we have investigated the SUSY contribution to the muon
MDM by regarding all the parameters in MSSM as free parameters.
Especially when $\tan\beta$ is large, the SUSY contribution is
enhanced, and some parameter region of MSSM is excluded not to
conflict with the present constraint on the muon MDM. Furthermore,
even in the case where $\tan\beta$ is not so large ($\tan\beta\lesssim
10$), $\Delta a_\mu^{\rm SUSY}$ may become comparable to the present
limit on the muon MDM, if the masses of the superparticles are quite
light (see Fig.~\ref{fig:now}a).

In MSSM, large $\tan\beta$ scenario is an interesting issue that has
been received attentions in recent years. One of the motivations of
large $\tan\beta$ is the unification of the masses of bottom and tau
in SUSY GUT~\cite{b-tau}. That is, in SUSY GUT where the Yukawa
coupling constants for bottom, $y_b$, and tau, $y_\tau$, are unified
at the GUT scale, the Yukawa coupling constant of bottom (or top) is
claimed to be significantly large in order to have the observed value
of the bottom mass. Thus, for the successful unification of $y_b$ and
$y_\tau$, large value of $\tan\beta$ is preferred. (Another solution
is to assume $\tan\beta\sim 1$ so that the Yukawa coupling constant
for top, $y_t$, becomes large.)  SUSY GUT based on SO(10) may give us
another motivation of large $\tan\beta$~\cite{so10}. In a simple
SO(10) GUT, all the Yukawa coupling constants (especially, $y_b$ and
$y_t$) are unified at the GUT scale. In this case, $\tan\beta$ as
large as $m_t/m_b\sim 50$ is required in order to make the hierarchy
between the top and bottom masses.  Furthermore, in some model in
which the masses of the light fermions are generated radiatively, we
need large value of $\tan\beta$~\cite{radiative_fermion_mass}. The new
Brookhaven E821 experiment will be a powerful test for such types of
large $\tan\beta$ scenarios.

Due to the fact that the SUSY contribution to the muon MDM strongly
depends on $\tan\beta$, we may be able to use the muon MDM for the
determination of $\tan\beta$, especially for the large $\tan\beta$
case. That is, by future experiments, in particular by NLC, we will be
able to measure the masses of the superparticles accurately, and it
can hopefully fix most of the parameters on which the muon MDM
depends.  Then, precise measurement of the muon MDM will give us an
useful information about $\tan\beta$.

Comparison of our results with those based on minimal
supergravity~\cite{9507386} may be interesting. In both cases, the
SUSY contribution to the muon MDM may become $O(10^{-8}-10^{-9})$ if
$\tan\beta$ is large. However, in our result, we can see several
interesting behaviors which hardly occur in the case of minimal
supergravity. That is, if we go away from the assumption of the
universal scalar mass, a cancellation may occur among several diagrams
when the mass splitting of left- and right-handed smuon is large.
Furthermore, in the case where the SUSY invariant Higgs mass $\mu_H$
is quite larger than the SUSY breaking parameters, diagram (N1) in
Fig.~\ref{fig:feyn_2} becomes significant, resulting in the
enhancement of $\Delta a_\mu^{\rm SUSY}$.

Finally, we would like to comment on the case with the flavor mixing
in the slepton mass matrices. In particular, even in the case of
minimal supergravity, the sfermion mass matrices receive
renormalization effects from the physics much above the electroweak
scale, such as the right-handed neutrino multiplets~\cite{lfv_neu,lfv}
or GUT~\cite{lfv_gut}, resulting in non-vanishing off-diagonal
elements of the slepton mass matrix.  If the off-diagonal elements of
the slepton mass matrices are substantially large, all the sleptons
contribute to the muon MDM, as we mentioned before. However, for the
case where the flavor mixing exists only in left- or right-handed
lepton mass matrix, the previous arguments are almost unchanged. If
both left- and right-handed slepton mass matrices have large
off-diagonal elements, situation changes.  Especially, in this case,
Yukawa coupling constant of tau can contribute to the muon MDM through
the Feynman diagram like (N1) in Fig.~\ref{fig:feyn_2}, and hence the
muon MDM may be enhanced.

Detailed analysis of this case is quite complicated since the muon MDM
depends on a large number of parameters. Thus, we only discuss the
case where the diagonal element of the left- and right-handed
sleptons, $\hat{m}_{L}^2$ and $\hat{m}_{R}^2$, are proportional to
unit matrix; $\hat{m}_{L,ii}^2=m_L^2$, $\hat{m}_{R,ii}^2=m_R^2$ ($i$:
not summed). First, we consider the case where one of $\hat{m}_{L}^2$
or $\hat{m}_{R}^2$ has off-diagonal element. In this case, the results
of the previous analysis are almost unaffected. For example, even if
$\hat{m}_{L,23}^2/\hat{m}_{L,22}^2=0.5$ (or
$\hat{m}_{R,23}^2/\hat{m}_{R,22}^2=0.5$), the correction to $\Delta
a_\mu^{\rm SUSY}$ is less than $\sim 10\%$. If both $\hat{m}_{L}^2$
and $\hat{m}_{R}^2$ have large off-diagonal elements, $\Delta
a_\mu^{\rm SUSY}$ may receive a large correction. Numerically, when
$\hat{m}_{L,23}^2/\hat{m}_{L,22}^2\sim\hat{m}_{R,23}^2/\hat{m}_{R,22}^2\sim
0.2$, the correction is $O(10\%)$. The correction gets larger as the
off-diagonal elements increase.

The new Brookhaven E821 experiment will give a strong impact on SUSY
models. By the experiment, the muon MDM is expected to be measured
with accuracy about $0.4 \times10^{-9}$. Furthermore, the uncertainty
in the theoretical prediction, which mainly comes from hadronic
contributions, is hoped to be reduced by several experiments like
VEPP-2M, DA$\Phi$NE and so on.  On the contrary, we may have the SUSY
contribution to the muon MDM to be of order $O(10^{-9})$ even if all
the superparticles are heavier than, say, 300GeV (see
Fig.~\ref{fig:r300gev}b) in which case we cannot detect the
superparticles even by NLC with $\sqrt{s}=500{\rm GeV}$.  Therefore,
we may be able to have a signal of the superparticles by using the
muon MDM even if the superparticles are out of the reach of the
forthcoming high energy colliders.

\section*{Acknowledgement}
The author would like to thank J.~Hisano and H.~Murayama for useful
discussions, and C.D.~Carone for reading of the manuscript.  This work
was supported by the Director, Office of Energy Research, Office of
High Energy and Nuclear Physics, Division of High Energy Physics of
the U.S. Department of Energy under Contract DE-AC03-76SF00098.

\appendix

\section{Functions $I_N$ and $J_N$}
\label{ap:i&j}

In this appendix, we show some useful formulae for the functions $I_N$
and $J_N$, which are defined as
\begin{eqnarray}
I_N(m_1^2,\cdots,m_N^2) &=&
\int\frac{d^4k}{(2\pi)^4i}
\frac{1}{(k^2-m_1^2)\cdots (k^2-m_N^2)},
\label{ap_I_N} \\
J_N(m_1^2,\cdots,m_N^2) &=&
\int\frac{d^4k}{(2\pi)^4i}
\frac{k^2}{(k^2-m_1^2)\cdots (k^2-m_N^2)}.
\label{ap_J_N}
\end{eqnarray}
The signs of the functions $I_N$ and $J_N$ are given by
\begin{eqnarray}
(-1)^{N} I_N(m_1^2,\cdots,m_N^2) &>& 0,
\\
(-1)^{N+1} J_N(m_1^2,\cdots,m_N^2) &>& 0,
\end{eqnarray}

The functions $I_N$ and $I_{N-1}$ are related as
\begin{eqnarray}
I_N(m_1^2,\cdots,m_N^2) = \frac{1}{m_1^2-m_N^2}
\{ I_{N-1}(m_1^2,\cdots,m_{N-1}^2) - I_{N-1}(m_2^2,\cdots,m_N^2) \},
\label{IN>=3}
\end{eqnarray}
and the explicit form of $I_2$ is given by
\begin{eqnarray}
I_2(m_1^2,m_2^2) = -\frac{1}{16\pi^2} 
\left\{ \frac{m_1^2}{m_1^2-m_2^2} \ln\left( \frac{m_1^2}{\Lambda^2} \right)
+ \frac{m_2^2}{m_2^2-m_1^2} \ln\left( \frac{m_2^2}{\Lambda^2} \right)
\right\}.
\label{I_2}
\end{eqnarray}
Notice that the function $I_2$ is logarithmically divergent, and hence
$I_2$ defined in eq.(\ref{I_2}) depends on a cut-off parameter
$\Lambda$. However, $I_N$ ($N\geq 3$) which is iteratively defined by
using eq.(\ref{IN>=3}) is independent of $\Lambda$, as it should be.
In addition, $J_N$ is related to $I_N$ and $I_{N-1}$ as
\begin{eqnarray}
J_N(m_1^2,\cdots,m_N^2) = I_{N-1}(m_1^2,\cdots,m_{N-1}^2)
+ m_N^2 I_N(m_1^2,\cdots,m_N^2).
\end{eqnarray}

In the case where all the masses $m_1$ -- $m_N$ are almost degenerate,
it is convenient to use the Taylar expansion of $I_N$. Define
\begin{eqnarray}
\epsilon_i \equiv \frac{\bar{m}^2-m_i^2}{\bar{m}^2}~~~(i=1-N),
\end{eqnarray}
with $\bar{m}$ being an arbitrary mass scale, then $I_N$ is expanded as
\begin{eqnarray}
I_N(m_1^2,\cdots,m_N^2) &=& \frac{(-1)^N}{16\pi^2}
\frac{1}{\bar{m}^{2(N-2)}} \sum_{p=0}^\infty
\frac{1}{(N+p-2)(N+p-1)}
\nonumber \\ &&
\times \sum_{j_1+\cdots +j_N=p} 
\epsilon_1^{j_1}\cdots\epsilon_N^{j_N}~~~(N\geq 3),
\end{eqnarray}
and for $N=2$, 
\begin{eqnarray}
I_2(m_1^2,m_2^2) = - \frac{1}{16\pi^2}
\left\{ \ln\left( \frac{\bar{m}^2}{\Lambda^2}\right) +1 \right\}
+ \frac{1}{16\pi^2}
\sum_{p=1}^\infty
\frac{1}{p(p+1)}
\sum_{j_1+j_2=p} 
\epsilon_1^{j_1}\epsilon_2^{j_2}.
\label{Taylar_I2}
\end{eqnarray}
Notice that eqs.(\ref{IN>=3}) -- (\ref{Taylar_I2}) are useful for
numerical calculations.

Furthermore, the function $I_N$ has mass dimension $(4-2N)$.
Therefore, we obtain
\begin{eqnarray}
\frac{d}{d\lambda}\left\{
\lambda^{2-N} I_N(\lambda m_1^2,\cdots,\lambda m_N^2)  \right\}
= 0,
\end{eqnarray}
which reduces to 
\begin{eqnarray}
(2-N) I_N(m_1^2,\cdots,m_N^2) + 
\sum_{i=1}^{N} m_i^2
I_{N+1} (m_1^2,\cdots,m_i^2,m_i^2,\cdots,m_N^2) =0.
\end{eqnarray}
Similar formula can be obtained for $J_N$;
\begin{eqnarray}
(3-N) J_N(m_1^2,\cdots,m_N^2) + 
\sum_{i=1}^{N} m_i^2
J_{N+1} (m_1^2,\cdots,m_i^2,m_i^2,\cdots,m_N^2) =0.
\end{eqnarray}

\newpage
\hspace{-\parindent}
{\Large\bf Figure caption}
\begin{figure}[h]
\caption
{Feynman diagrams which give rise to the muon MDM in the mass
eigenstate basis. The external lines represent the muon (straight)
and the photon (wavy).}
\label{fig:feyn_1}
\caption
{Feynman diagrams which give rise to the muon MDM in the mass
insertion method.}
\label{fig:feyn_2}
\caption
{The SUSY contribution to the muon MDM, $\Delta a_{\mu}^{\rm SUSY}$,
in the $\mu_H$-$m_{G2}$ plane. The right-handed smuon mass is taken to
be $m_{\tilde{\mu}R}=100{\rm GeV}$. We take $\tan\beta =30$, and the
left-handed smuon mass $m_{\tilde{\mu}L}$ is (a) 100GeV, (b) 300GeV
and (c) 500GeV. The numbers given in the figures represent the value
of $\Delta a_\mu^{\rm SUSY}$ in units of $10^{-9}$.}
\label{fig:r100gev}
\caption
{Same as Fig.~\protect{\ref{fig:r100gev}} except for
$m_{\tilde{\mu}R}=300{\rm GeV}$.}
\label{fig:r300gev}
\caption
{Same as Fig.~\protect{\ref{fig:r100gev}} except for
$m_{\tilde{\mu}R}=1{\rm TeV}$.}
\label{fig:r1tev}
\caption
{The SUSY contribution to the muon MDM, $\Delta a_{\mu}^{\rm SUSY}$,
in the $\mu_H$-$m_{G2}$ plane. The right-handed smuon mass
$m_{\tilde{\mu}R}$ is determined so that $\Delta a_{\mu}^{\rm SUSY}$
takes its minimal value. We take $\tan\beta =30$, and the left-handed
smuon mass $m_{\tilde{\mu}L}$ is taken to be (a) 100GeV, (b) 200GeV,
(c) 300GeV and (d) 500GeV. The numbers given in the figures represent
the value of $\Delta a_\mu^{\rm SUSY}$ in units of $10^{-9}$.}
\label{fig:min}
\caption
{Constraints on the $\mu_H$-$m_{G2}$ plane for $\tan\beta =30$ from
the negative searches for the neutralinos and the charginos. The
numbers on the figure represent the lowerbound on the chargino mass in
units of GeV. The contour with $m_{G2}\leq 45{\rm GeV}$ corresponds to
the constraint from LEP~\protect{\cite{lep,PLB350-109}}.}
\label{fig:lep}
\caption
{Contours which gives the threshold value, {\it i.e.} $\Delta
a_\mu^{\rm SUSY}=-9.0\times 10^{-9}$ (dotted line) and $\Delta
a_\mu^{\rm SUSY}=19.0\times 10^{-9}$ (solid line). The right-handed
smuon mass $m_{\tilde{\mu}R}$ is determined so that $|\Delta
a_{\mu}^{\rm SUSY}-5.0\times 10^{-9}|$ is minimized.  The values shown
in the figures represent those of $\tan\beta$, and we take the
left-handed smuon mass $m_{\tilde{\mu}L}$ to be (a) 100GeV, (b) 200GeV
and (c) 300GeV.}
\label{fig:now}
\end{figure}

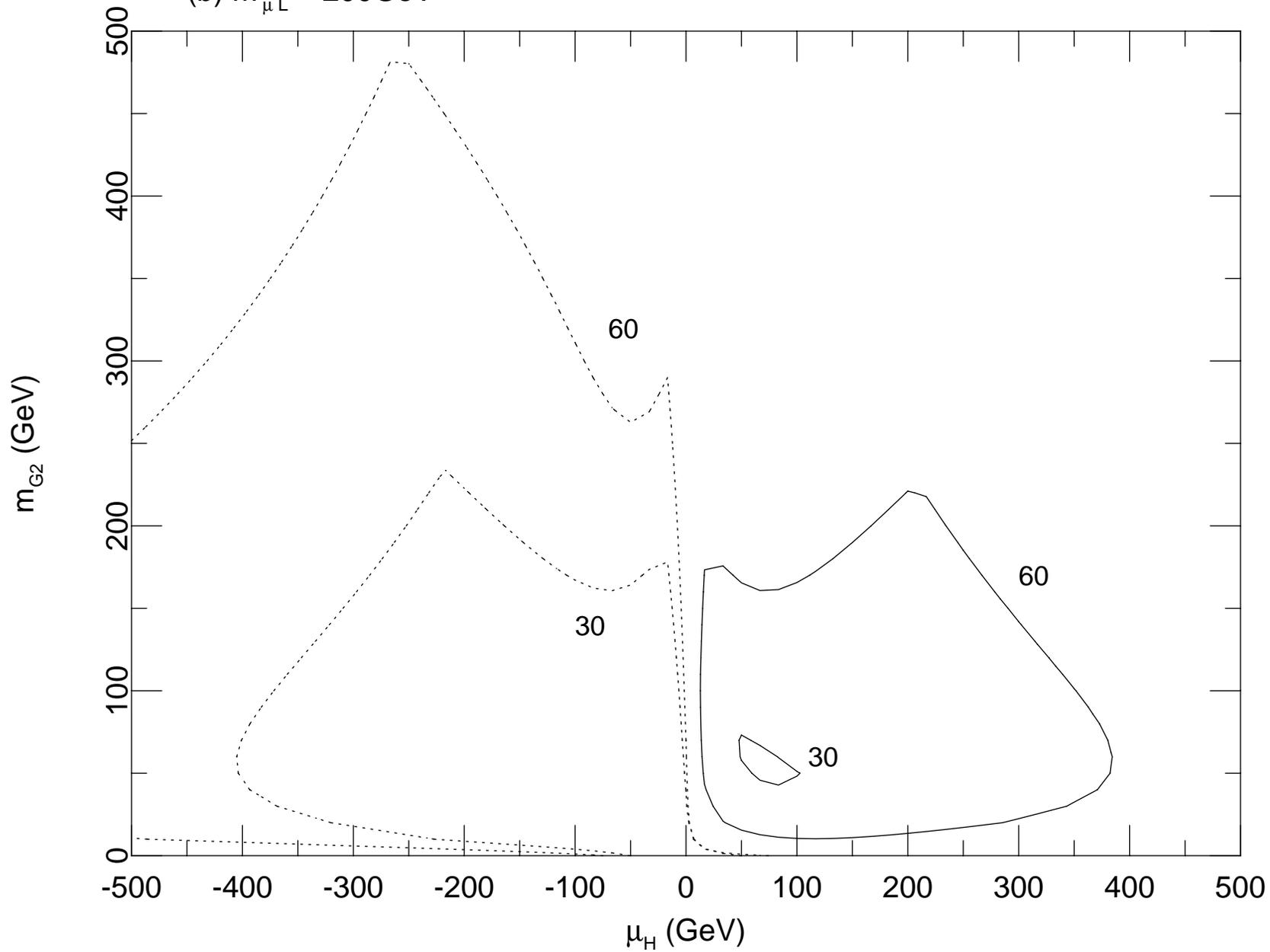
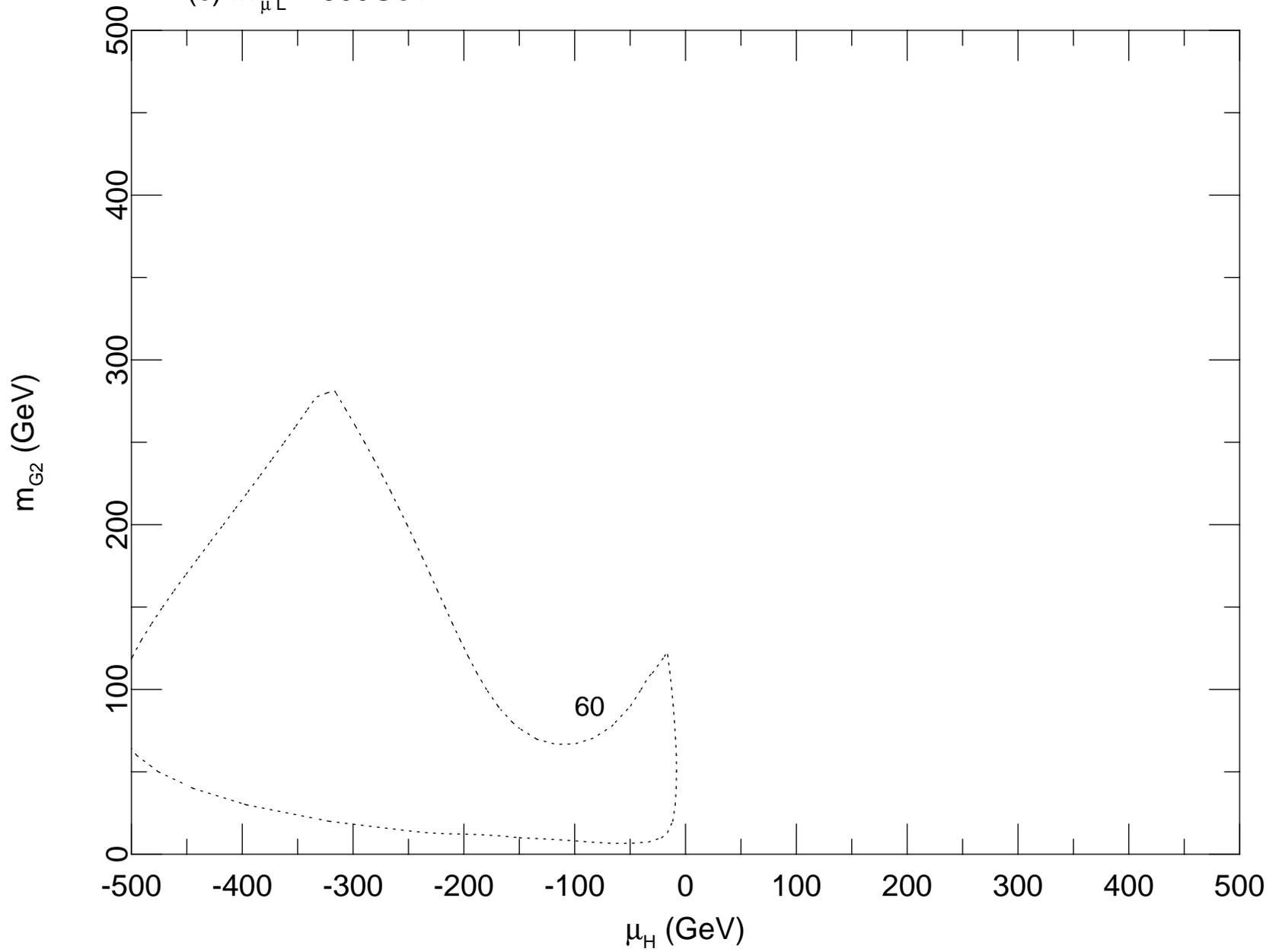
\end{document}